\documentclass[epjCONF]{svjour}
\usepackage{graphics}
\usepackage[varg]{txfonts} 
\usepackage[latin1]{inputenc}
\session-title{MESON2012 - 12th International Workshop on Meson Production, Properties and Interaction}
\begin{document}
\title{Search for $\eta\rightarrow e^+e^-$ decay with the WASA experiment}
\author{M.~Ber{\l}owski for WASA-at-COSY Collaboration\inst{1}\fnmsep\thanks{\email{Marcin.Berlowski@fuw.edu.pl}}}
\institute{[1] National Centre for Nuclear Research, Warsaw, Poland}
\abstract{
Nowadays the field of searching for a new physics became a very interesting subject in a light meson decays due to
a recent results from KTeV collaboration which found the $3.3\sigma$ disagreement between Standard Model theory and
their results of $\pi^\circ\rightarrow e^+e^-$ branching ratio measurement \cite{AD07}. They propose to explain
this discrepancy with a new $U$ boson particle that interacts both with meson and virtual photon producing $e^+e^-$
pair \cite{YK07}. The same effect could be observed in eta meson decay into electron-positron. The current branching ratio limit
\cite{MB08,GA12} is far away from the predicted non-Standard Model theory and due to that fact it cannot distinguish
between Standard Model and more exotic explanation. The following report shows the analysis highlights for searching
for a such effect in $pp\rightarrow pp(\eta\rightarrow e^+e^-)$ at 1.4 GeV produced in WASA@COSY experiment.
}
\maketitle
\section{Introduction}\label{introduction}
\begin{figure}[htbp]
\centering
\resizebox{0.35\columnwidth}{!}{\includegraphics{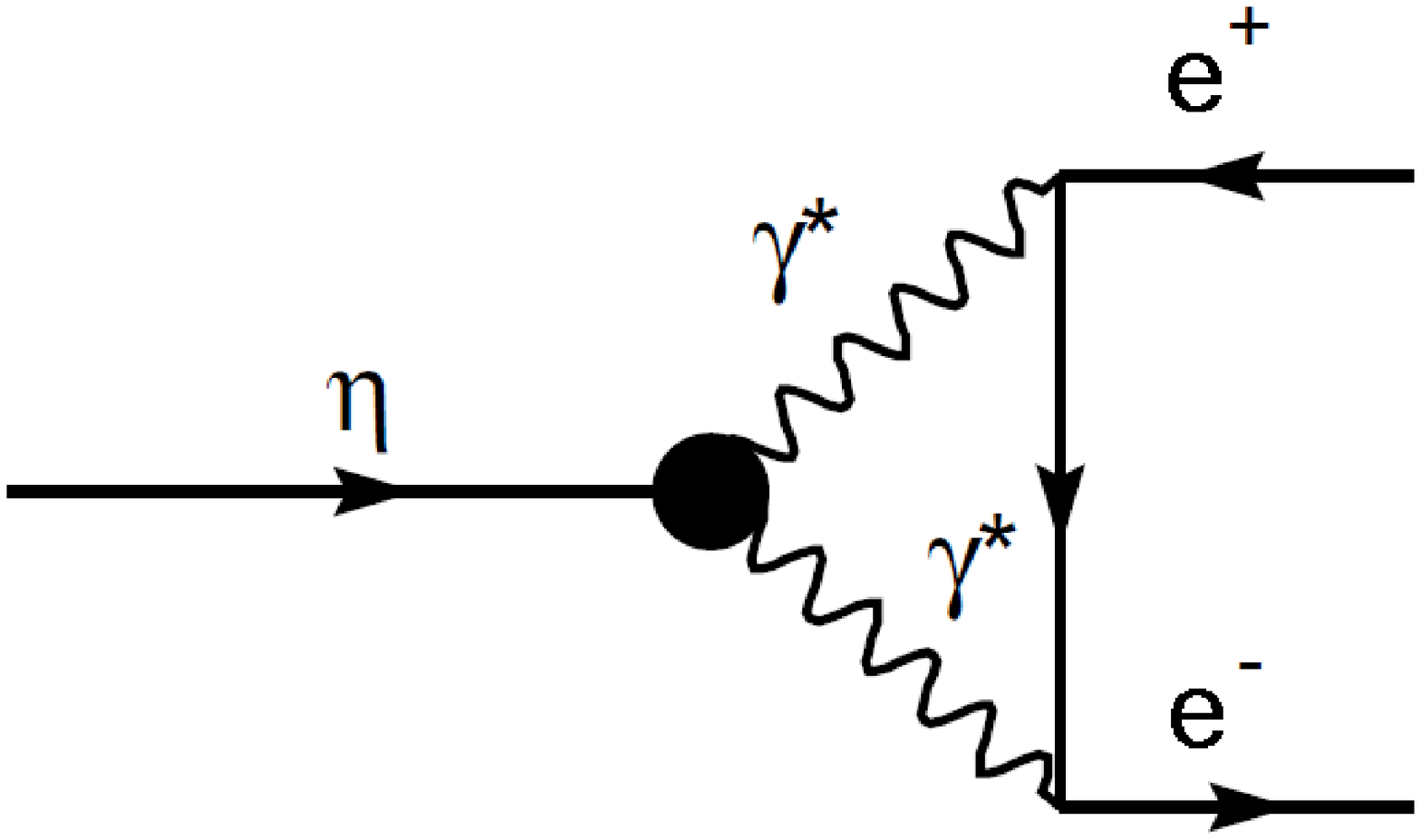}}
\resizebox{0.35\columnwidth}{!}{\includegraphics{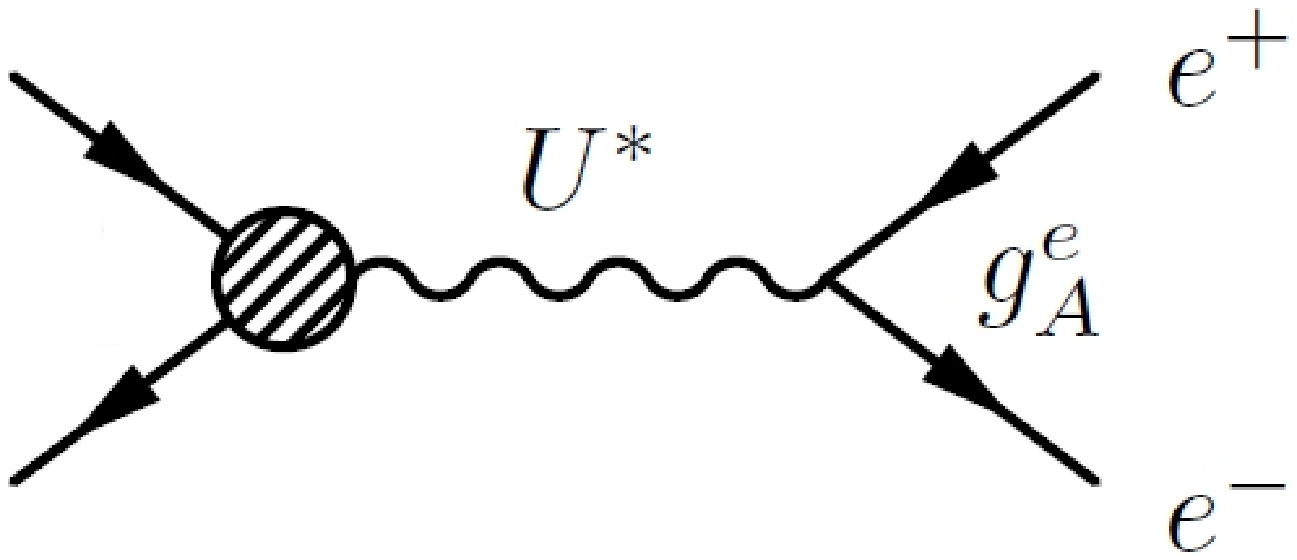}}
\caption{The dominating conventional mechanism for $\eta\rightarrow e^+e^-$ decay (left panel), an example of process with intermediate U boson interaction (right panel).}
\label{diagrams}
\end{figure}

Due to the fact that very low branching ratio (BR) is expected in the Standard Model, $\eta\rightarrow e^+e^-$ decay is very hard to observe.
In the Standard Model, the decay proceeds dominantly through the electromagnetic interaction (see fig.\ref{diagrams} left)

and it is suppressed relative to $\eta\rightarrow\gamma\gamma$ by $\alpha^2$ and by $(m_e/m_\eta)^2$ from helicity conservation:
\begin{center}
$BR[\eta\rightarrow e^+e^-]\sim BR[\eta\rightarrow\gamma\gamma]\cdot\alpha^2\cdot(m_e/m_\eta)^2$
\end{center}
The upper limit for this decay $BR_{exp}<2.7\times10^{-5}\ at\ CL=90\%$ comes from CELSIUS/WASA experiment
\cite{MB08} and the most recent value comes from HADES collaboration and sets the limit to $<5.6\times10^{-6}~at~CL=90\%$ \cite{GA12}. This numbers are at least four orders of magnitude larger then value predicted from the Standard Model calculations
($BR_{theo}\sim10^{-9}$). The small probability of this fourth-order electromagnetic transition makes the decay sensitive
to hypothetical interactions that arise from physics beyond Standard Model \cite{YK07,PF06,QC08,LB82} (see fig.\ref{diagrams} right). An observation of a signal above $\sim10^{-9}$ level could be evidence for an unconventional process which enhances this decay rate.

\section{Experiment}\label{experiment}
The 2 weeks experiment took place at Institute for Nuclear Physics of the Forschungszentrum Juelich in Germany. For this purpose WASA
detection system installed at the COSY storage ring was used. The proton beam of energy 1.4 GeV (2.14 GeV/c in momentum) 
was scattered on frozen hydrogen pellets crossing the beam line.
The reaction products along with scattered protons were detected and measured in WASA detector (protons in the Forward part and meson decay products in the Central part).
The detailed description of the WASA detector can be found in \cite{HA04}.
Runs were dedicated for the eta meson decays coming from $pp\rightarrow pp\eta$ reaction. For this analysis a special trigger system was used demanding a high energy deposit in each of Central Detectors halves. This type of trigger should treat equally the following decays of the eta meson: $\eta\rightarrow\gamma\gamma$, $\eta\rightarrow e^+e^-\gamma$, $\eta\rightarrow e^+e^-e^+e^-$, $\eta\rightarrow e^+e^-$.

\section{Analysis}\label{analysis}
\begin{figure}[htbp]
\centering
\resizebox{0.4\columnwidth}{!}{\includegraphics{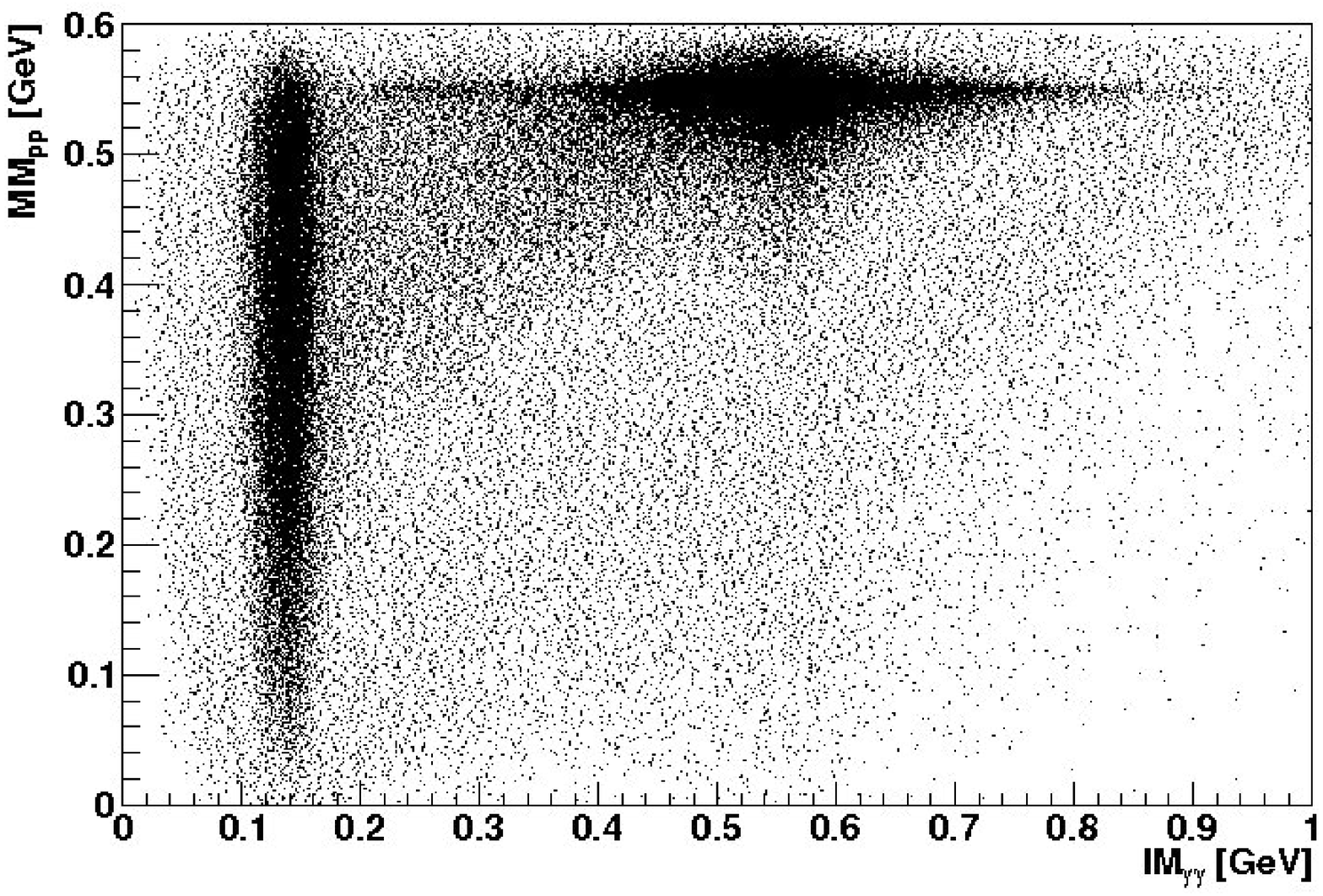}}
\resizebox{0.4\columnwidth}{!}{\includegraphics{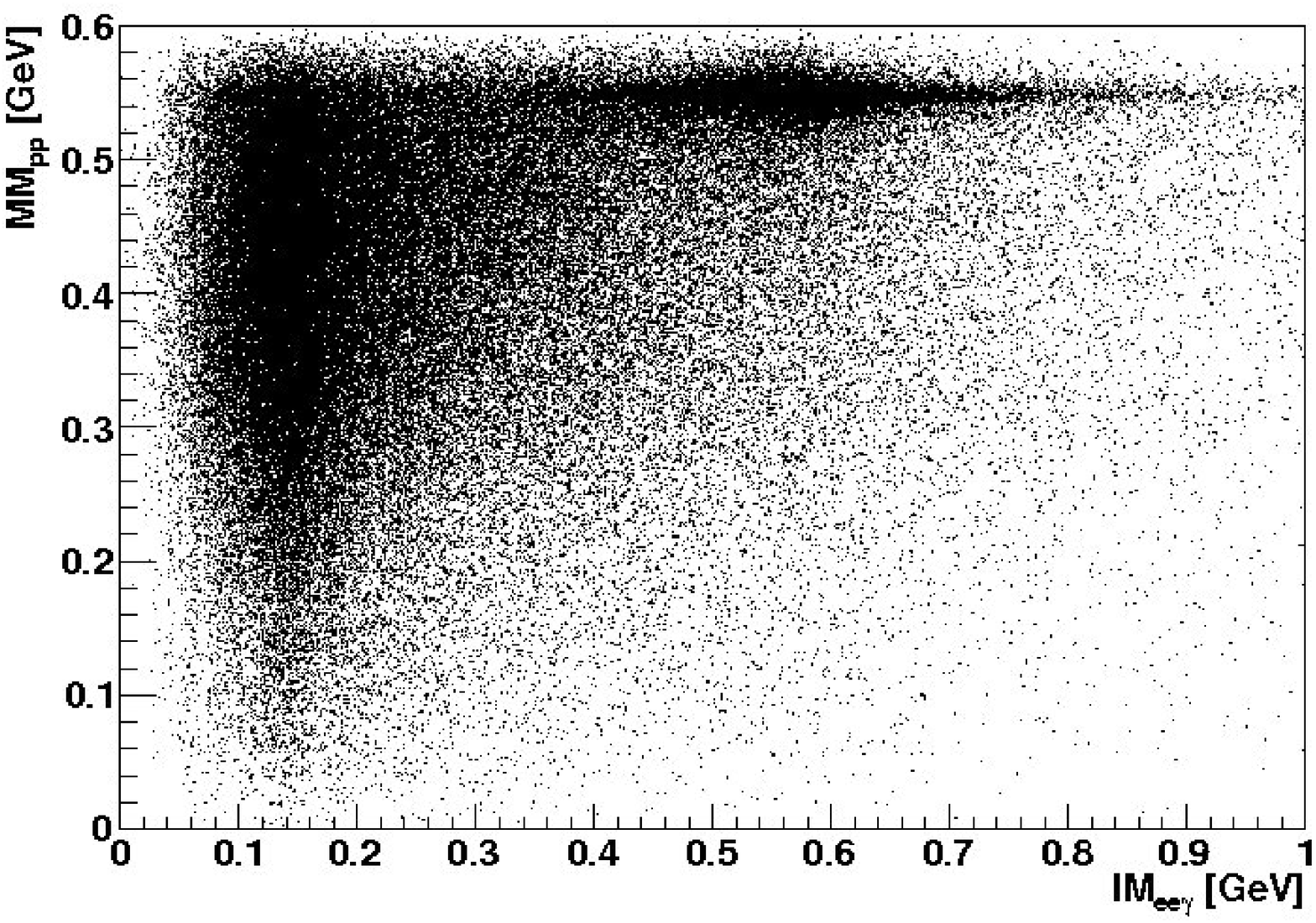}}
\caption{Missing mass of two protons versus invariant mass of two photons (left panel), Missing mass of two protons versus invariant mass of $e^+e^-\gamma$ (right panel).}
\label{etaggeeg}
\end{figure}

In order to determine the number of eta mesons produced, the data sample of $\eta\rightarrow\gamma\gamma$ decays were used. From the number of $\eta\rightarrow\gamma\gamma$ events reconstructed (see fig.\ref{etaggeeg} left), knowing the BR for the decay one was able to estimate this number to $N_{eta}=\sim4.4\times10^7$. As a crosscheck to this normalization procedure $\eta\rightarrow e^+e^-\gamma$ decay channel was used (see fig.\ref{etaggeeg} right). The second channel also provided essential information about detector response to electron-positron pairs of various energies. Both channels served also as an experimental field for studying the performance of the trigger system.

\begin{figure}[htbp]
\centering
\resizebox{0.4\columnwidth}{!}{\includegraphics{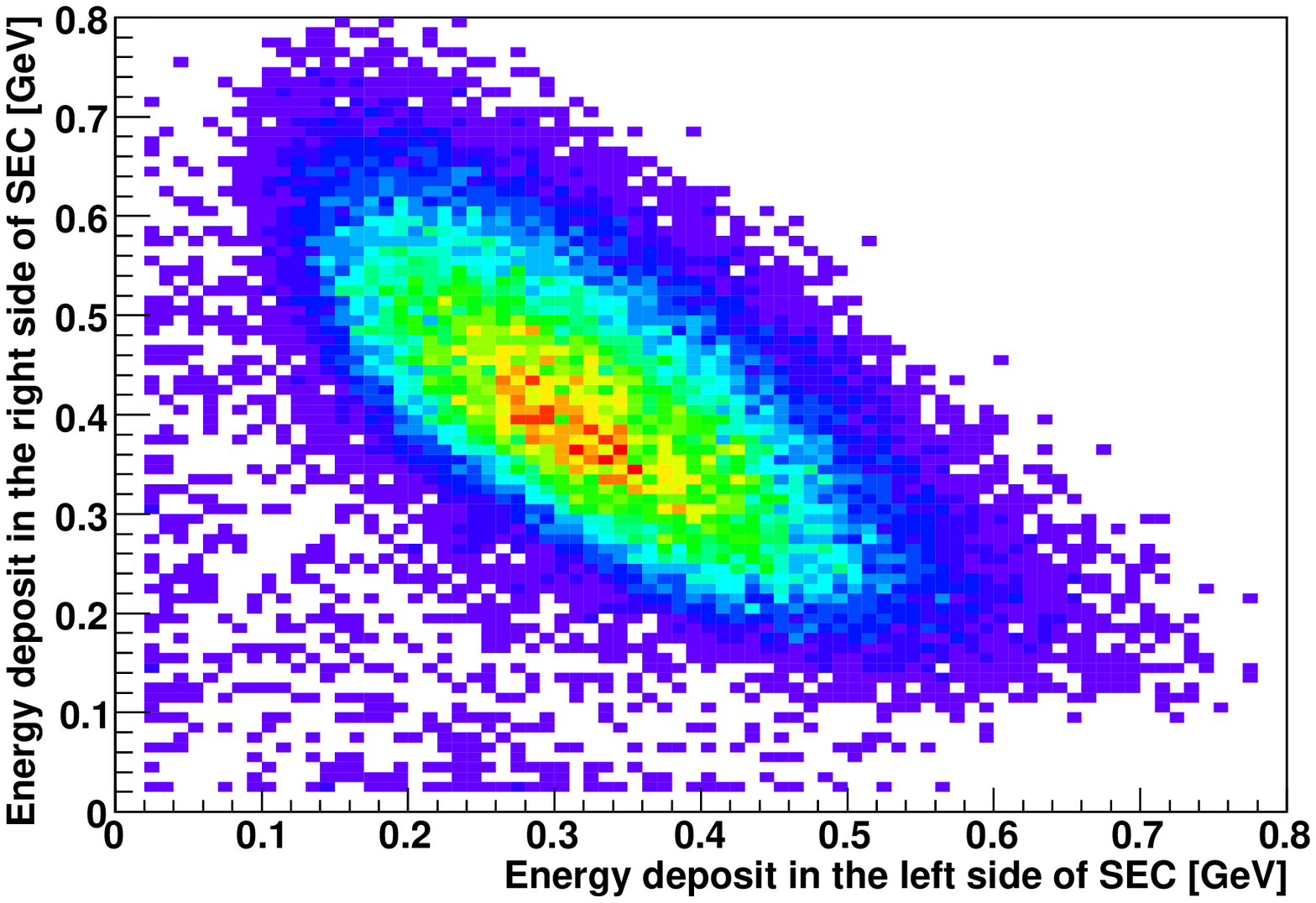}}
\resizebox{0.4\columnwidth}{!}{\includegraphics{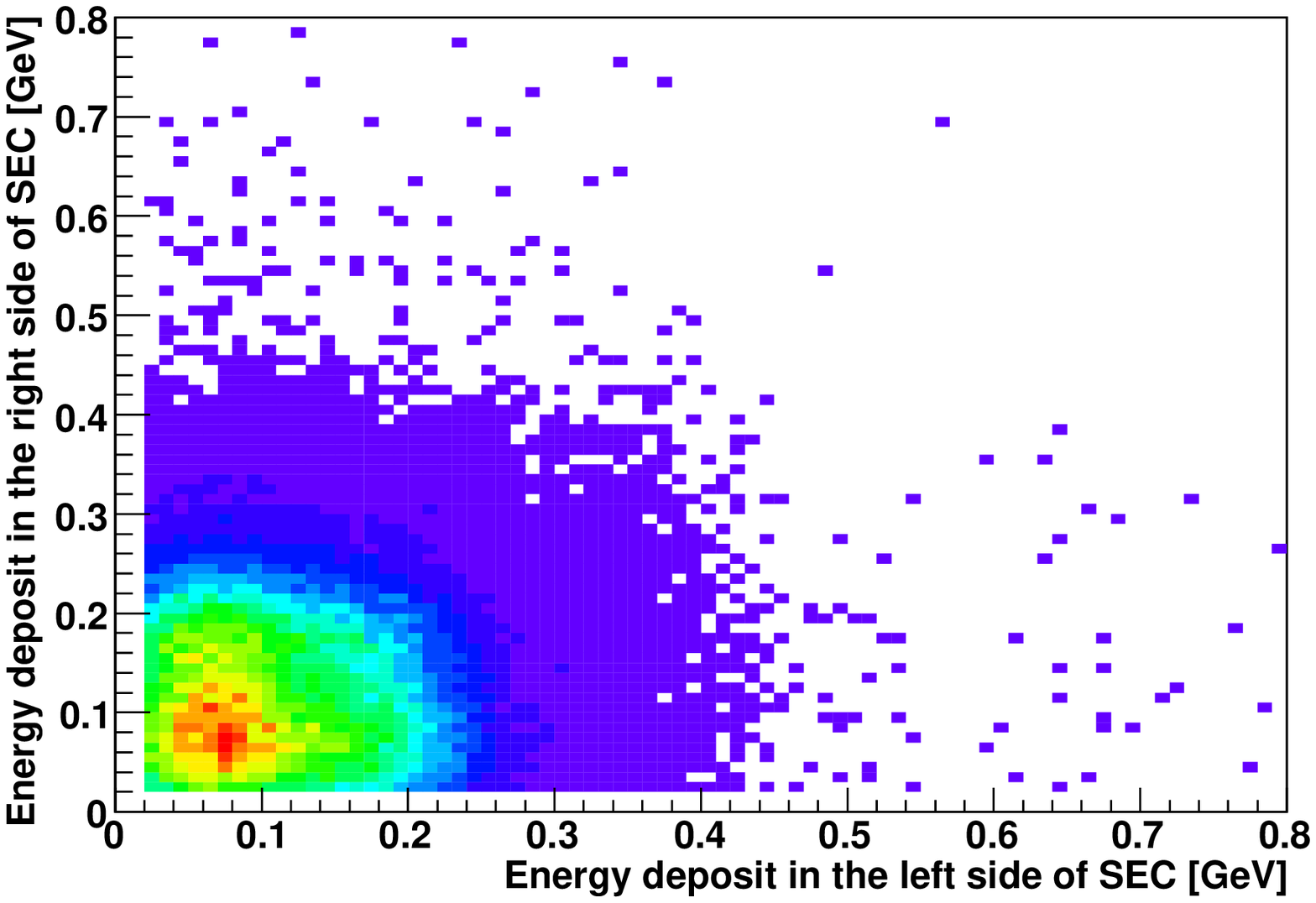}}
\caption{Energy deposits in both sides of the energy calorimeter for Monte Carlo simulated $\eta\rightarrow e^+e^-$ signal (left panel) and Monte Carlo direct production of two charged pions (right panel).}
\label{edeps}
\end{figure}

Particle selection for the $\eta\rightarrow e^+e^-$ included demanding at least two protons in the Forward Detector (if there were more, two closest in time were used), veto on neutral particles in the Central Detector of energy more than 20 MeV, at least two oppositely charged (if more, two giving the greatest opening angle were used) and time correlation between both charged in Central Detector and protons in the Forward Detector.
As for $\eta\rightarrow e^+e^-$ the first step was determination of the main background sources which appeared to be: $pp\rightarrow pp\pi^+\pi^-$, $pp\rightarrow pp(\eta\rightarrow e^+e^-\gamma)$ and $pp\rightarrow p(\Delta\rightarrow p[\gamma^*\rightarrow e^+e^-])$. The direct two charged pion production has 100 time larger cross-section then for the eta meson production and the same number of charged particles in the final state. The single Dalitz decay of the eta meson has the same final state particles (coming from the same as in $\eta\rightarrow e^+e^-$ initial stage), if the mass of virtual photon is large enough and real photon is not observed. The radiative decay of Delta(1232) resonance has the same final state as $\eta\rightarrow e^+e^-$. According to the Monte Carlo distributions for the channels above and signal simulation optimization procedure for cuts were evoked. The particles in the Central Detector were identified in two different methods: energy deposited in electromagnetic calorimeter (see fig.\ref{edeps} and also by using ratio of deposited energy to particle momentum.

\begin{figure}[htbp]
\centering
\resizebox{0.35\columnwidth}{!}{\includegraphics{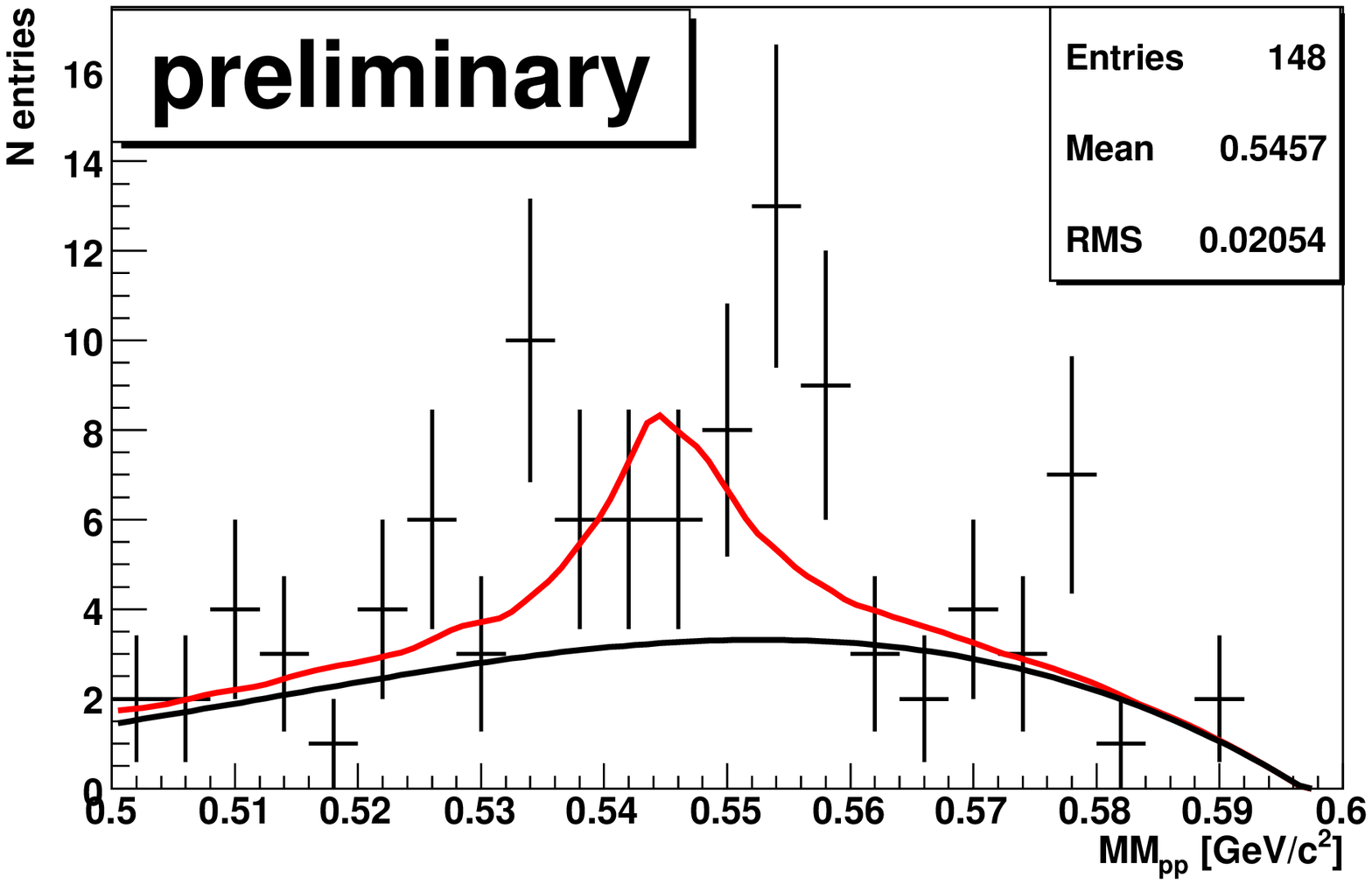}}
\resizebox{0.35\columnwidth}{!}{\includegraphics{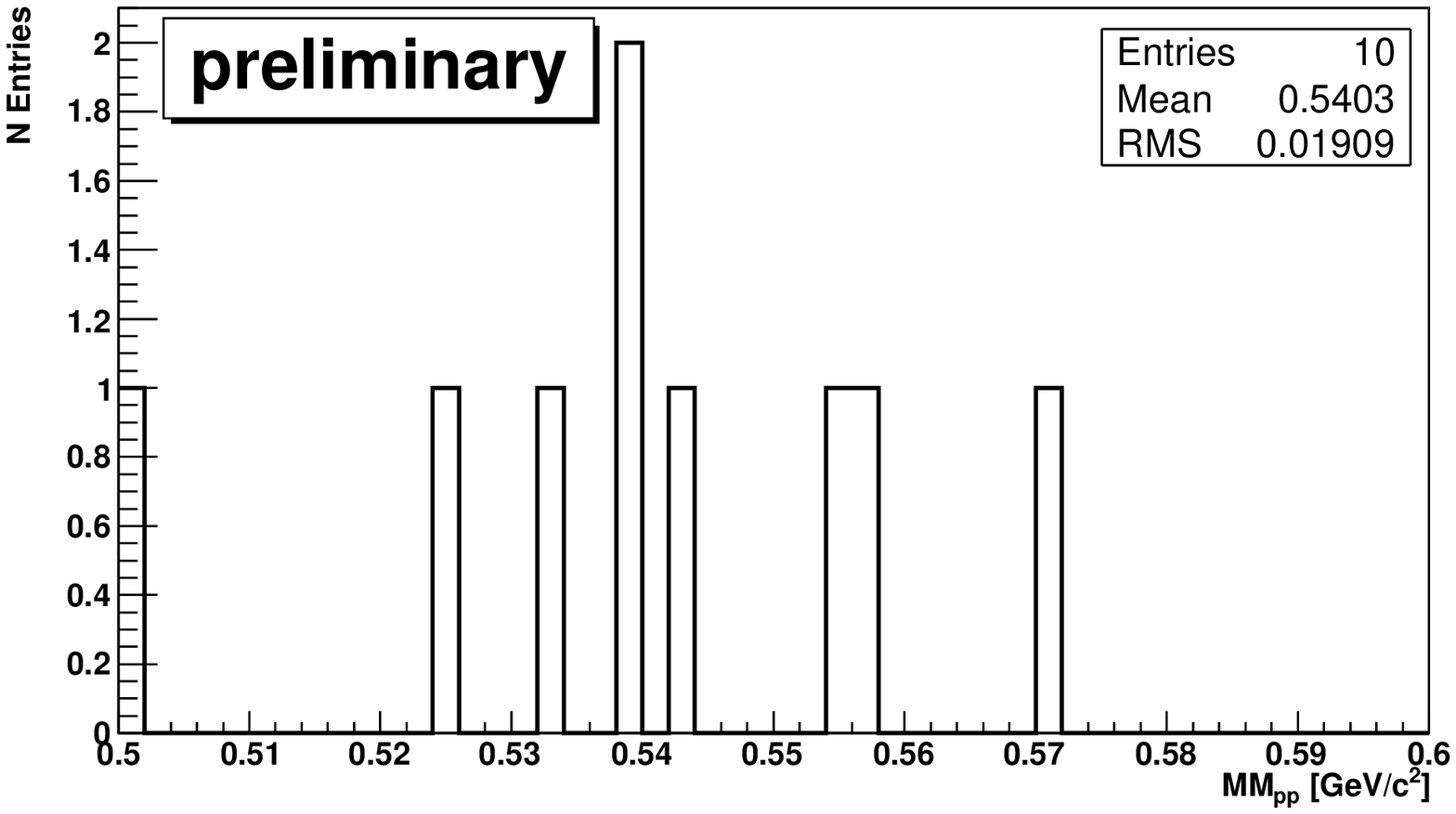}}
\caption{Plot of the missing mass of two protons. Black curve - polynomial fit to the background, red curve - the shape of expected Monte Carlo $\eta\rightarrow e^+e^-$ signal plus polynomial background (left panel). The final distribution of the missing mass of two protons (right panel).}
\label{data}
\end{figure}

After the analysis we are left with 148 event candidates in whole data sample on the plot of missing mass of two protons using cut optimization, obtaining Monte Carlo signal acceptance = $5.5\%$. If in addition we use reduction of background coming from $pp\rightarrow pp\pi^+\pi^-$ and $pp\rightarrow p(\Delta\rightarrow p[\gamma^*\rightarrow e^+e^-])$ with the polynomial fit subtraction and the remaining expected number of events coming from $pp\rightarrow pp(\eta\rightarrow e^+e^-\gamma)$ background = $4.6\pm1.5$ events (see fig.\ref{data} left). After additional cuts (on total missing mass, momentum and energy and other) the possible signal is consistent within errors with the expected number of events coming from $\eta\rightarrow e^+e^-\gamma$ (see fig.\ref{data} right). The lack of the signal events leads us to the preliminary BR limit:
\begin{center}
$BR_{limit}=4.6\times10^{-6}\ at\ CL\ 90\%$ (preliminary),
\end{center}
which is an order of magnitude below present limit \cite{MB08} for branching ratio of $\eta\rightarrow e^+e^-$ decay and at the same level as the most recent measurement from HADES Collaboration \cite{GA12}. Eight times larger statistic of the same reaction with similar trigger was recently collected and analysis has started.

\section{Acknowledgements}
We acknowledge support by the Polish Ministry of Science and Higher Education under grant No. 86/2/N-DFG/07/2011/0.

\end{document}